\documentclass[journal = jacsat, manuscript=article]{achemso}
\usepackage{graphicx}
\usepackage{color}
\usepackage{caption}
\usepackage{amssymb}
\usepackage{dirtytalk}
\usepackage{gensymb}
\usepackage{booktabs}
\usepackage{subfigure}
\usepackage{mathtools}
\usepackage{threeparttable}
\usepackage{tikz}
\usetikzlibrary{shapes.geometric, arrows}

\usepackage{longtable}

\author{Zeyu Zhang}
\author{Dylan Valente}
\author{Yuliang Shi}
\author{Dil K. Limbu}
\author{Mohammad R. Momeni$^*$$^{\dagger}$}
\author{Farnaz A. Shakib}
\email{momeni@njit.edu,shakib@njit.edu}

\affiliation{Department of Chemistry and Environmental Science, New Jersey Institute of Technology, Newark 07102, NJ United States}

\title{EC-MOF/Phase-I: A computationally ready database of electrically conductive metal-organic frameworks with high-throughput structural and electronic properties}

\begin{document}

\begin{abstract}
The advent of $\pi$-stacked layered metal-organic frameworks (MOFs) opened up new horizons for designing compact MOF-based devices as they offer unique electrical conductivity on top of permanent porosity and exceptionally high surface area. By taking advantage of the modular nature of these electrically conductive (EC) MOFs, an unlimited number of materials can be created for applications in electronic devices such as battery electrodes, supercapacitors, and spintronics. Permutation of structural building blocks including different metal nodes and organic linkers results in new systems with unprecedented and unexplored physical and chemical properties. With the ultimate goal of providing a platform for accelerated materials design and discovery, here, we lay the foundations towards creation of the first comprehensive database of EC-MOFs with an experimentally guided approach. The first phase of this database, coined EC-MOF/Phase-I, is comprised of 1,061 bulk and mono-layer structures built by all possible combinations of experimentally reported organic linkers, functional groups and metal nodes. A high-throughput screening (HTS) workflow is constructed to implement density functional theory calculations with periodic boundary conditions to optimize the structures and calculate some of their most significantly relevant properties. Since research and development in the area of EC-MOFs has long been suffering from the lack of appropriate initial crystal structures, all the geometries and property data have been made available for the use of the community through the online platform that is developed in the course of this work. This database provides comprehensive physical and chemical data of EC-MOFs as well as convenience of selecting appropriate materials for specific applications, thus, accelerating design and discovery of EC-MOF-based compact devices.
\end{abstract}

\noindent\section{Introduction\label{sec1}}
In the last two decades, metal-organic frameworks (MOFs) have constituted one of the fastest growing fields in chemistry and materials science{\cite{kitagawa2022metal}} with a wide range of applications in adsorption,{\cite{morris2008gas}} separation{\cite{li2012metal}} and catalysis{\cite{lee2009metal}}, to name a few.{\cite{furukawa2013chemistry}} MOFs are a class of porous crystalline materials obtained through a process usually referred to as reticular synthesis.{\cite{yaghi2019reticular}} Selected metal nodes and organic linkers, also called secondary building units (SBUs), are connected via strong coordination bonds to form ordered and permanently porous architectures.{\cite{yaghi2003reticular}} The modular nature of MOFs provides many opportunities to tailor their physical and chemical characteristics, which has led to more than 110,000 MOFs reported to date according to the Cambridge Structural Database (CSD).{\cite{groom2016cambridge}} Traditional MOFs are mostly classified as insulators with wide band gaps, which limits their further utilization in electrical and optical devices.{\cite{wang2021two}} The discovery of $\pi$-stacked layered MOFs, also referred to as two-dimensional (2D) MOFs, in 2012 opened a new research direction in this area due to the remarkable electrical conductivity of these materials compared to traditional MOFs.{\cite{hmadeh2012new}} As a result, more 2D MOFs are being reported by researchers emphasizing on the excellent electrical conductivity and magnetic properties introducing them as viable candidates for field-effect transistors, {\cite{wu2017porous}} supercapacitors,{\cite{li2017conductive}} superconductors,{\cite{huang2018superconductivity}} spintronics{\cite{dong2018coronene}} and cathode materials in different metal-ion batteries.{\cite{wada2018multielectron}} Normally, electrically-conductive (EC) MOFs contain ortho-substituted organic linkers coordinated to transition metal nodes, forming extended $\pi$-conjugated 2D sheets. Weak van der Waals interactions allow stacking of these 2D sheets to form bulk crystalline materials with one-dimensional channels in the stacking direction. Layers of synthesized EC-MOFs to date generally contain extended $\pi$-conjugated organic linkers and (usually) early 3d transition metal nodes as building blocks, providing the necessary paths for charge transport along both in-plane and out-of-plane directions.{\cite{zhang2021metal}} Various building blocks for EC-MOFs are reported in the literature and as mentioned above, MOFs, in general, can be rationally designed by choosing different combinations of organic linkers and metal node building blocks. However, considering the vast and virtually infinite chemical space of MOFs, it is extremely labor intensive and time consuming to synthesize all different combinations of building blocks to find the best materials for any desired application. A more efficient and systematic way is to create a comprehensive database of different classes of MOFs and then screen them for desired applications using high-throughput screening (HTS) techniques.{\cite{rosen2019identifying}} Chung {\textit{et al.}}{\cite{chung2014computation}} created a computation-ready, experimental (CoRE) MOF database with over 5,000 MOFs based on CSD in 2014. Various other datasets have evolved from the CoRE MOF database, such as CoRE MOF 2014+DDEC{\cite{nazarian2016comprehensive}} where partial atomic charges were determined for 50\% of the reported MOFs using density functional theory (DFT) calculations  as well as CoRE MOF 2014-DFT-optimized{\cite{nazarian2017large}} where DFT geometric relaxation was performed for 879 structures. CoRE MOF database itself was updated in 2019 with the total number of MOFs being increased to 14,000.{\cite{chung2019advances}} On the other hand, the Cambridge Crystallographic Data Centre (CCDC) created a MOF subset based on existing CSD which contains the largest number of experimentally synthesized MOFs to date.{\cite{moghadam2017development}} A total number of 69,666 MOFs were gathered in this subset after screening the original CSD based on 7 different criteria and removing solvent molecules from the MOF pores. A more recent work by Rosen {\textit{et al.}}{\cite{rosen2021machine}} in 2021 introduced a new database called Quantum MOF (QMOF) database containing 15,713 MOFs that were successfully optimized and analyzed by HTS periodic DFT workflows. 

\begin{figure*}[t]
\centering
\includegraphics[width=0.99\linewidth]{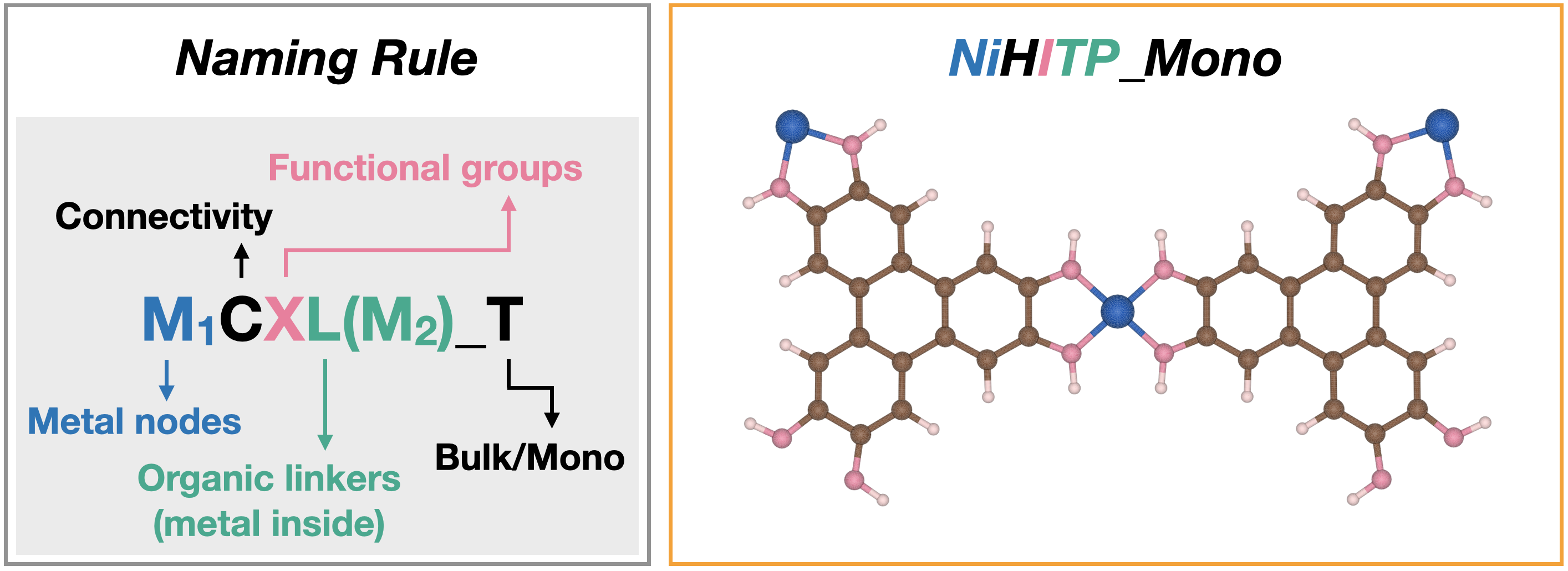}
\caption{Naming rules used for the structures included in EC-MOF/Phase-I database (left) and an example of a built NiHITP crystal structure (right).}
\label{FigureSubset2}
\end{figure*}

Not only experimental MOFs are collected in different databases, a hypothetical MOF database is also recently built by Wilmer {\textit{et al.}}{\cite{wilmer2012large}} where 137,953 new MOFs were created from a combination of 102 different building blocks. More than 300 candidates were selected with excellent methane storage capacity. These carefully prepared databases allow selection of materials with desired properties and performance for specific applications by fast screening of hundreds and/or thousands of structures.
However, they all exclude or at best partially include $\pi$-stacked layered EC-MOFs since they are a very new class of materials. Considering the wide potentials of EC-MOFs, it is crucial to first build a comprehensive database for them which will then allow various HTS techniques to be routinely applied in order to accelerate materials design and discovery. Here, we report the first installation of our experimentally-guided, computationally-ready database of EC-MOFs, coined EC-MOF/Phase-I, containing 1,061 bulk and mono-layer structures. This database is available to the community use via the developed online platform during the course of this study at https://ec-mof.njit.edu. All the structures in this database follow a comprehensive naming rule as shown generally and with an example in Figure~\ref{FigureSubset2}. 
This naming rule, which will be fully explained in the next section, gives enough information to the user about the nature of metal nodes, functional groups, organic linkers, connectivity between building blocks and type of unit cells. In this way, the users can easily have access to the desired structure in the database through the choice of each of these components where they can build and download their structure. Furthermore, the users not only have access to the crystal structures but also geometric data and electronic properties obtained using our in-house HTS workflow. We will provide the details about the applied procedure and developed software for building the database in the next section of this manuscript. In section 3 we will present the results of our HTS investigation of the created 1,061 systems. Section 4 outlines future directions for this research while concluding remarks will be made in section 5.

\noindent\section{Computational Developments\label{sec1}}
\subsection{EC-MOFs from literature} 
\begin{figure*}[t]
\centering
\includegraphics[width=0.99\linewidth]{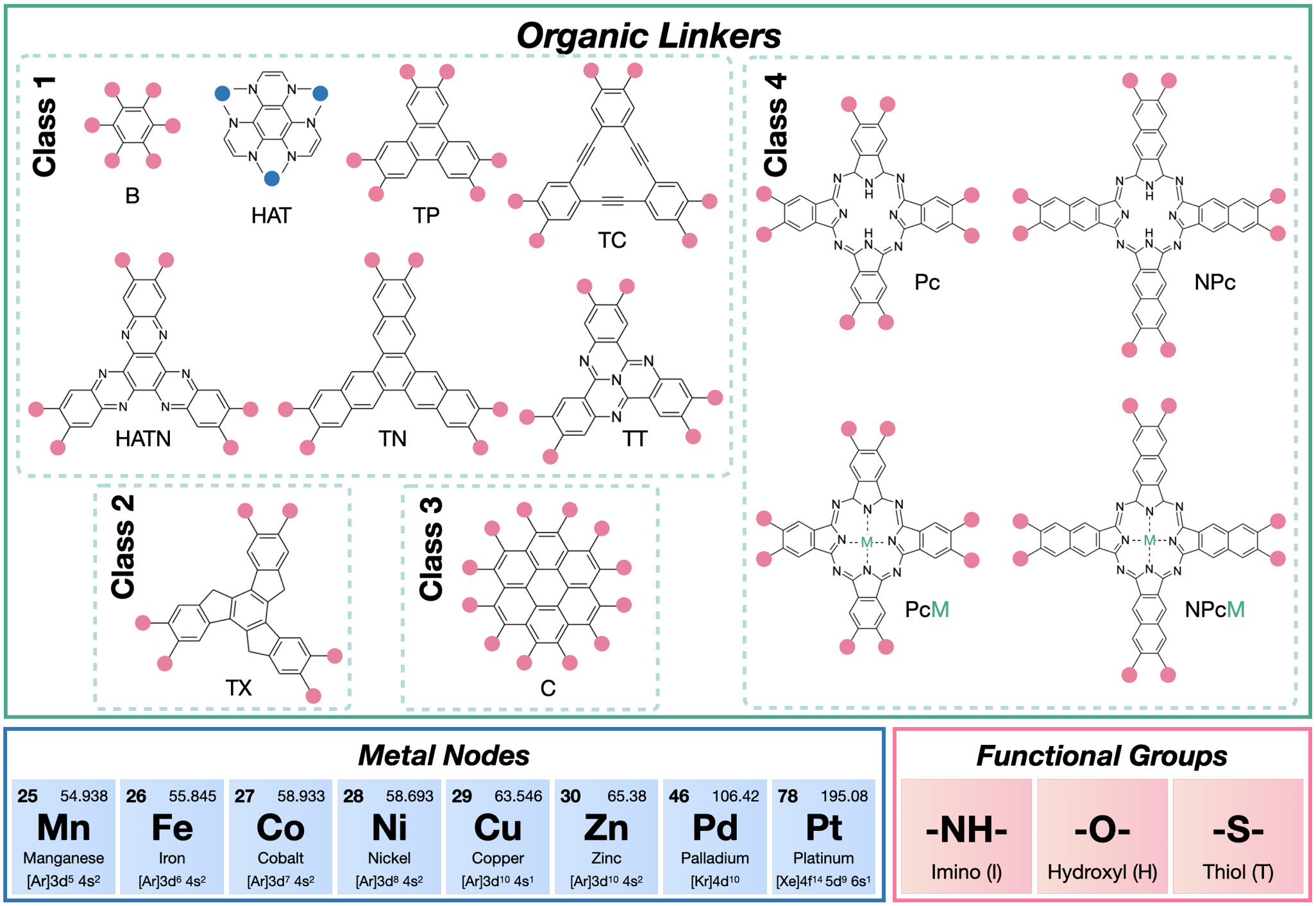}
\caption{Three subsets of structural building components including organic linkers, metal nodes, and functional groups (green, blue and pink boxes). Connecting sites of organic linkers to metal and functional groups are highlighted by blue and red dots, respectively}
\label{FigureSubset1}
\end{figure*}

As mentioned above, reticular chemistry is defined as a process that simple molecular building blocks are linked by strong coordinative bonds to form extended crystalline architectures such as those of MOFs.{\cite{lyu2020digital, lyu2020digital, xu2020anisotropic}} 
In this work, we first performed a thorough literature survey and summarized all reports on EC-MOFs that have been either synthesized and/or theoretically investigated (see Supporting Information (SI) section S1 for details). With the intention of developing a structure creation tool for automatic generation of the initial crystal structures of EC-MOFs for our database, we initially focused on identifying structural features that induce the highest electrical conductivity. Accordingly, we restricted the first version of our database to $\pi$-stacked EC-MOFs with planar layers and extended $\pi$-conjugation through organic linkers and metal nodes with 2+ oxidation state allowing for effective $d-\pi$ conjugation. Notably, Hofmann-type MOFs are excluded from the first version of our database because depending on the nature of the coordinative bonds they can form 3D structures where $\pi$-stacking can occur along the in-plane rather than out-of-plane direction. Moreover, presence of Pt$^{2+}$ nodes can disturb in-plane electrical conductivity in some of these MOFs.\cite{Sakaida:2016,Sakaida:2017} Other classes of MOFs were excluded on the basis of presence of twists in the layers\cite{Liu:2020} which can disrupt extended in-plane conjugation, presence of bi- and tri-nuclear metal nodes\cite{Liao:2018} which can disrupt effective in-plane $d-\pi$ conjugation, and more than 4 \AA~interlayer distance\cite{Foster:2016} that reduces out-of-plane $\pi-\pi$ interactions drastically. The subset of EC-MOFs gathered in our database follow the color-coded naming rule depicted in Figure~\ref{FigureSubset2} based on three structural building components including organic linkers, metal nodes, and functional groups. Figure~\ref{FigureSubset1} demonstrates all the structural components used in building EC-MOF/Phase-I database.
The eight metal nodes including Mn, Fe, Co, Ni, Cu, Zn, Pd and Pt, are identified by $M_1$ in the name while $X$ denotes 3 functional groups as $H$ for hydroxyl, $I$ for imino, and $T$ for thiol that connect to the metal nodes. The linker subset of structural building components, denoted as $L$ in our naming, consists of 10 organic linkers including benzene (B), triphenylene (TP), trinaphthylene (TN), truxene (TX), coronene (C), tribenzocyclyne (TC), tetraazanaphthotetraphene (TT), hexaazatriphenylene (HAT), hexaazatrinaphthalene (HATN), phthalocyanine (Pc) and naphthalocyanine (NPc). Pc and NPc linkers can accommodate one more transition metal atom inside. Transition metals including Fe, Co, Ni, Cu, Zn and Pd are observed in synthesized Pc and NPc-based MOFs, therefore, PcM and NPcM with 6 different interior metal centers are considered. The connections between metal nodes and organic linkers happen through 3, 4 or 6 bidentate sites. Accordingly, 6, 8 or 12 functional groups are placed around one linker. This is identified with $C$ (as in Connectivity) in the name of the EC-MOFs which can be replaced with $H$ for hexa, $O$ for octa and $P$ for per. Furthermore, $M_2$ is the metal atom inside organic linker, if existed. At last, $T$ indicates the type of unit cell which has two options, bulk or mono-layer. Naturally, some of the resulted structures are originated from reported works but most are hypothetical. Hence, our building strategy screens all combinations of the building blocks but keeps a tight connection between experimental and theoretical studies. 

\noindent\subsection{Crystal Structure Producer (CrySP)}
\begin{figure*}[t]
\centering
\includegraphics[width=0.99\linewidth]{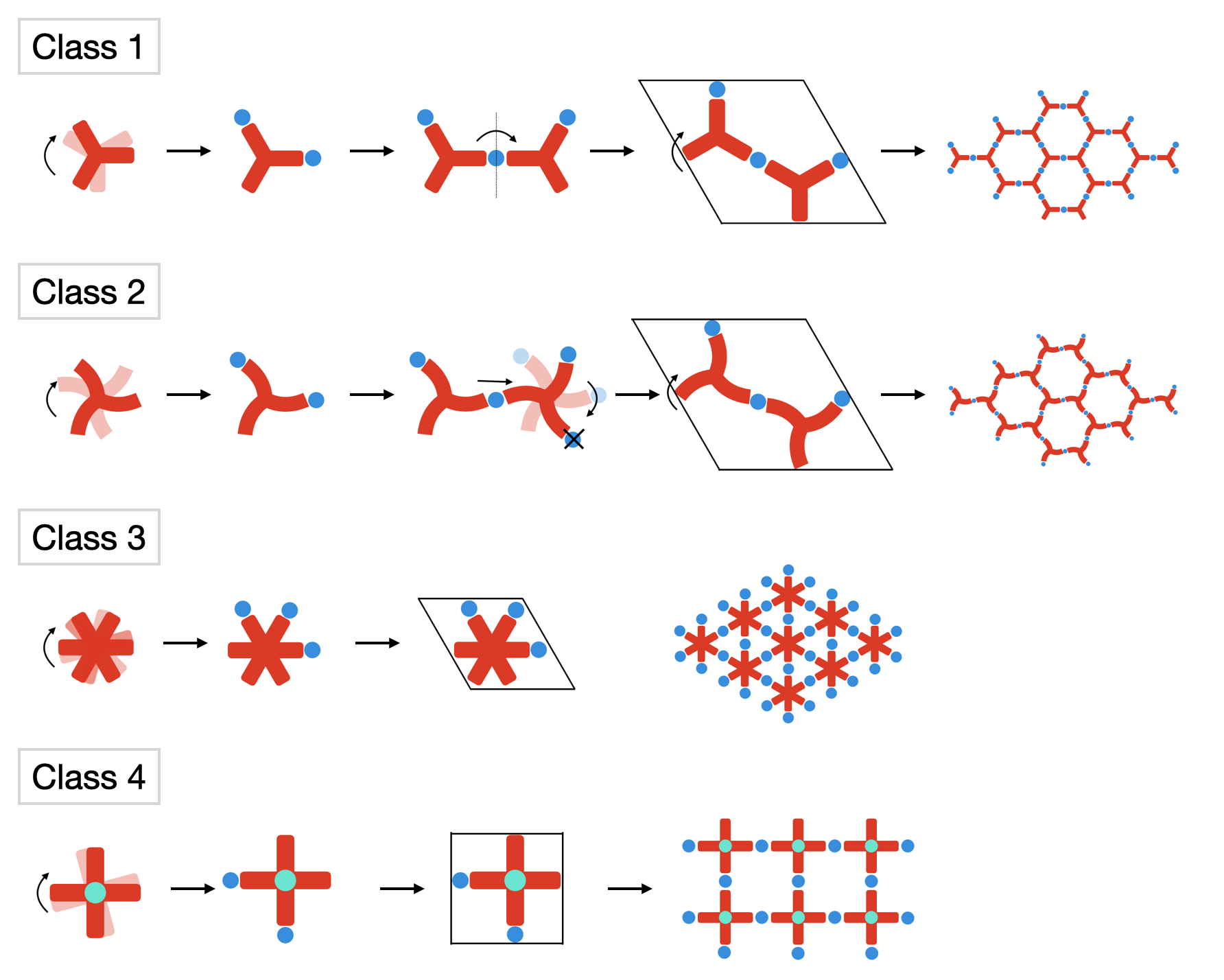}
\caption{Building logic of CrySP for four different classes of MOFs created in this work. Classification of organic linkers is given in Figure~\ref{FigureSubset1} with our formulations for calculation of cell vectors of different classes provided in the SI.}
\label{FigureClasses}
\end{figure*}

Based on the criteria established in the previous section, we identified four different classes of EC-MOFs according to the shape and connectivity of the organic linkers as demonstrated in Figure~\ref{FigureClasses}. An in-house structure building tool, coined Crystal Structure Producer (CrySP), is developed to create periodic structures of EC-MOFs that fit into these four geometric classes. The CrySP algorithm starts by rotating the organic linker to the desired position, then functional groups are placed around the organic linker according to the connectivity of the linker. Next, metal nodes are added according to the bond length between metals and functional groups which is specified in advance. After necessary transformations, the structure is rotated to fit into the specified unit cell. Structures in our Database, coined EC-MOF/Phase-I, can be classified into 3 different lattices, including honeycomb (hcb), hexagonal lattice (hxl) and square lattice (sql).\cite{liu2022conductive} 
We further divide hxl structures into 2 classes due to the unique building procedure needed for TX-based MOFs. The details of the building procedures of four classes may vary as schematically shown in Figure~\ref{FigureClasses}. In Class 1, CrySP calls up the xyz coordinates of the desired organic linker and places metal nodes around it in appropriate positions and distances. Then, the entire structure is reflected along the \textit{y} axis. The last steps are moving and rotating the structure to fit into the specified unit cell. In class 2, the reflecting step is replaced by moving and rotating operation due to the special symmetry of the linker. In class 3 and 4, once the organic linker is placed at the center of the unit cell, functional groups and metals are added around the organic linker without any transformational operation. At the same time, CrySP calculates cell parameters for each class of materials by taking the positions of metal nodes as reference points. Mathematical details can be found in the Supporting Information (SI), section S1 and Figure S1. Finally, CrySP creates the structures in the desired formats including XYZ, Crystallographic Information file (CIF) or POSCAR. The resulted structures at this stage are all mono-layers. To optimize the mono-layer structures a vacuum space of 20 \AA~is added to the \textit{c} direction of the unit cells. Bulk structures, containing 2 layers in the unit cell, are created with a fixed inter-layer distance along out-of-plane direction, i.e., 3.25~\AA, while considering different crystal packing known as AA or AB stacking \cite{ko2018conductive}. Accordingly, a total of 1,072 structures are created by CrySP and gathered in the first version of our EC-MOF database.

\noindent\subsection{Details of our high-throughput workflows and periodic electronic structure calculations}
\begin{figure}[h!]
\centering
\includegraphics[width=0.6\linewidth]{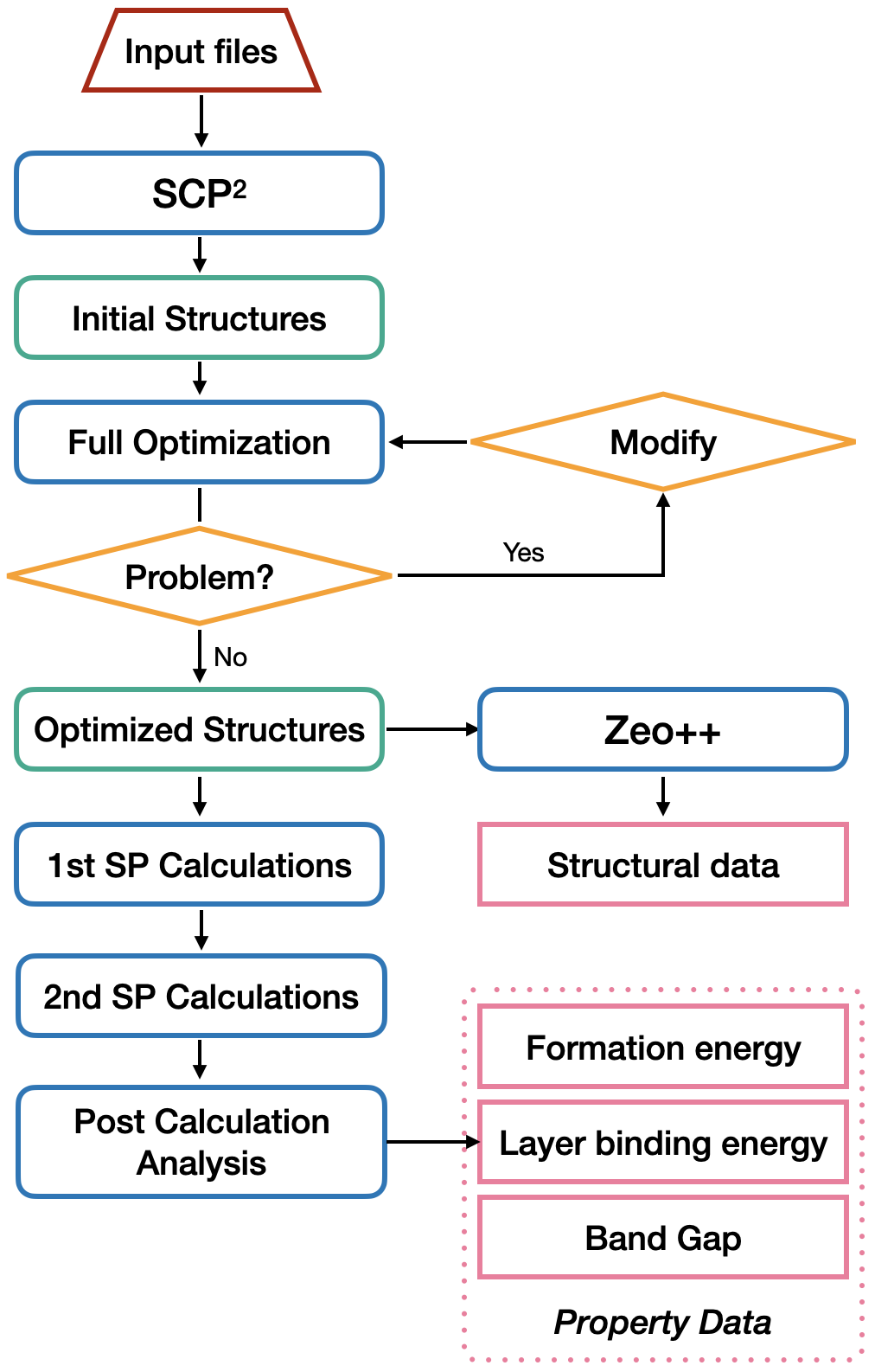}
\caption{High-throughput screening workflow including different steps of geometric optimizations and single-point energy calculations for all the structures contained in our EC-MOF database.}
\label{FigureWorkFlow}
\end{figure}

To maximize the advantages of EC-MOF/Phase-I, we apply high-throughput screening (HTS) techniques to explore different properties of these materials, as shown in Figure \ref{FigureWorkFlow}.
Multiple steps of density functional theory (DFT) calculations are performed in our HTS workflows, which are all carried out using Vienna \emph{ab} initio simulation package (VASP) version 5.4.4.{\cite{kresse1993,kresse1994,kresse1996,kresse19962}} Interactions between electrons and ions are described by projector-augmented wave (PAW) potentials{\cite{blochl1994projector,kresse1999ultrasoft}} with a cut-off energy of 500 eV. Spin-polarized calculations are performed for all systems. Perdew-Burke-Ernzenhof (PBE) functional with Grimme’s damped D3 dispersion correction within the generalized gradient approximation (GGA) formalism{\cite{perdew1996generalized,grimme2010consistent}} was employed in this work which provides accurate results for our EC-MOF systems, as discussed in greater details in our recent work.{\cite{zhang2021metal}} First stage of the workflow is to optimize all structures obtained from CrySP to find the ground state energy configuration for both ions and electrons. The criterion for optimization convergence is set to less than $10^{-4}$ eV for electronic energy and the magnitude of the largest force acting on the atoms is set to less than 0.02 eV/\AA. Three degrees of freedom are allowed to change, atomic positions, cell volume and cell shape for bulk systems. Cell volume is fixed during optimization of mono-layers because we use a slab model with c = 20 {\AA} to prevent interaction between layers from different periodic cells. If the optimization encounters structure and setting-related issues or has any difficulty to reach convergence, an automatic debugging process is executed to collect error messages which automatically updates the input files for re-optimizations. Once the optimization is completed, the 1$^{st}$ round of single-point (SP) calculations is carried out to check if there is any structural or setting-related issues and to provide a good initial electron density for the following calculations. The convergence criteria for electronic energy is set to $1\times10^{-5}$ eV. The Brillouin zones are sampled using $2\times2\times6$ or $2\times2\times4$ k-point mesh for bulk systems depending on different ratios of cell parameters and $3\times3\times1$ k-point mesh for all mono-layer systems. Gaussian smearing method, with a smearing width of 0.05 eV is adopted in this step to provide accurate electronic energy and density information. The 2$^{nd}$ round of SP calculations reads the electron density calculated in the 1$^{st}$ round, which increases efficiency. The smearing method is changed to the tetrahedron method with Blöchl corrections which provides more accurate results for the calculated band gaps and density of states (DOS).{\cite{sholl2011density}} DFT in its GGA formalism is known for underestimating band gaps. Hence, Hubbard U approach (DFT+U), which semi-empirically optimizes the Coulomb interaction potential (U), is adopted in this step to give a better description of electronic structures. Only $d$ and $f$ electrons were treated by this approach with the employed semi-empirical U parameters for each metal reported in the SI, Section 2 Table S1.

\begin{figure}[h!]
\centering
\includegraphics[width=0.8\linewidth]{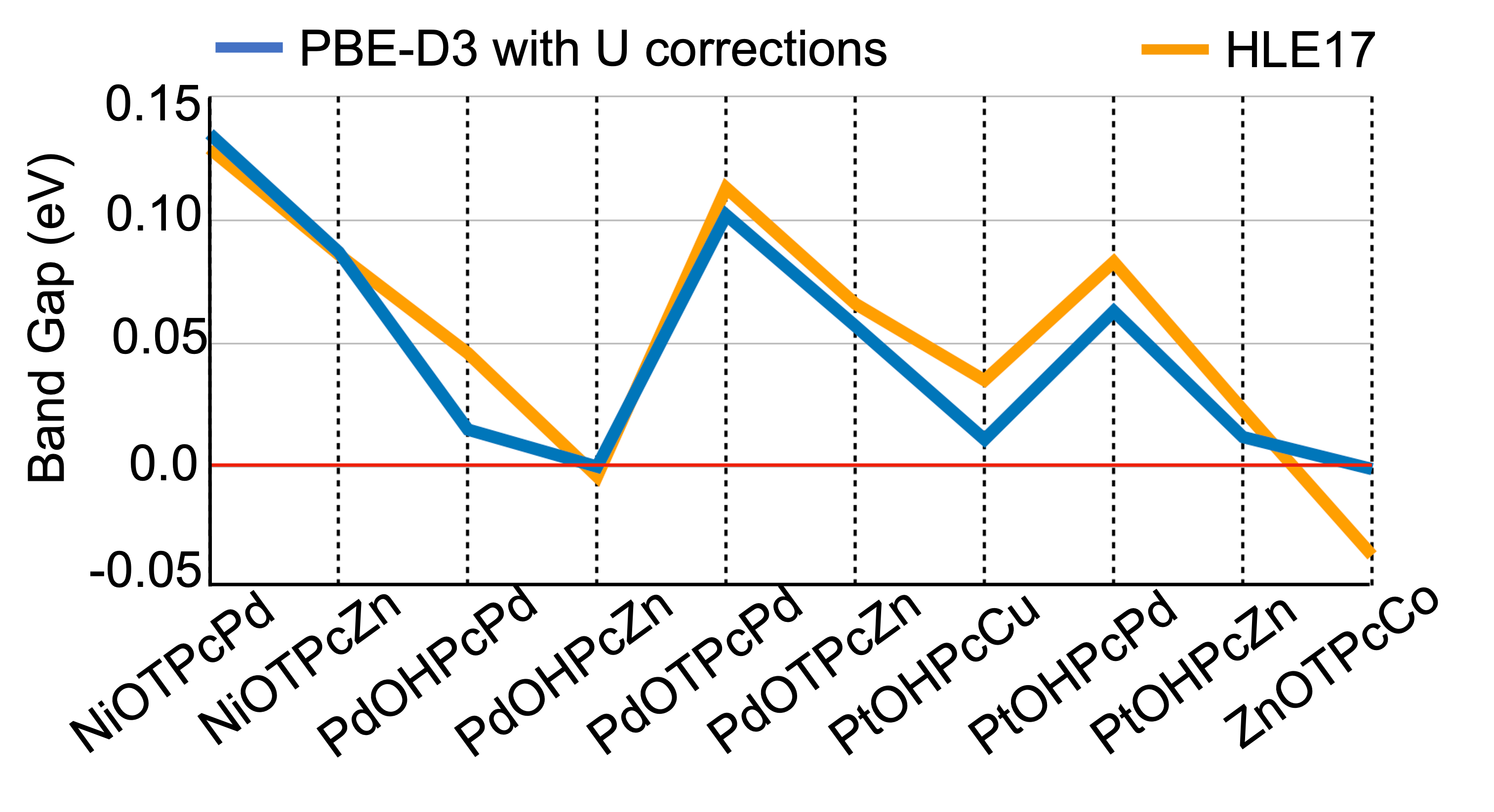}
\caption{Calculated band gaps of a subset of mono-layer structures from EC-MOF/Phase-I database. The results from DFT+U and HLE17 functionals are shown in blue and orange, respectively.}
\label{fig:HLE}
\end{figure}

To validate the accuracy of our calculated band gaps, a benchmark study was carried out comparing the band gap values of a subset of mono-layer structures from our EC-MOF database obtained from PBE-D3 with U correction to a meta-GGA functional, HLE17.\cite{verma2017hle17, choudhuri2019hle17} HLE17 is already shown to provide accurate band gap values of complex materials and is implemented in Minnesota VASP Functional Module codes. The results in Figure~\ref{fig:HLE} show a good agreement between DFT+U and HLE17 calculations. It is worth mentioning that not only the observed trend from DFT+U follows the same trend from HLE17 but also the absolute values of the computed DFT+U band gaps are close to the HLE17 results. Accordingly, successful completion of both SP steps provided necessary information for us to extract the final property data from our EC-MOF/Phase-I database.

\noindent\subsection{Post calculation analysis} Our HTS calculations provide us with different properties including geometric, energetic and electronic data for all structures generated and contained in our EC-MOF/Phase-I database. Periodic crystal structures are extracted as CIF files after the geometry optimization stage. Analysis of the Voronoi network, as implemented in Zeo++,\cite{zeo++} is used to determine different geometric data including largest cavity diameter (LCD), accessible void-volume fraction and accessible surface area, using a Helium probe with the radius of 1.4 {\AA}, for all bulk EC-MOFs. From SP calculations, the total energy and electronic structures of the systems are obtained. Absolute formation energy which is an important indicator of the stability of different structures, is calculated as:{\cite{kirklin2015open}}
\begin{equation}
E_{f}=E_{tot}-\frac{1}{N}\sum_{i=1}^{N}x_i\mu_i
\label{eq1}
\end{equation}
where $E_{tot}$ is the calculated energy for the entire bulk material, $N$ is the total number of atoms with $x_i$ and $\mu_i$ being the number and chemical potential of element $i$ in the structure. Calculations of each element in their most stable configuration are implemented to find the energy values per atom which is needed in the equation above. Calculated chemical potentials for elements are listed in the SI Table S2. Inter-layer binding energies were calculated as following:{\cite{huang2022band}}
\begin{equation}
 E_{b}=(E_{M}-{\frac{1}{n}}E_{B})/A  
 \label{eq2}
\end{equation}
where $E_{M}$ is the calculated energy of the mono-layer, $n$ is the number of layers in the bulk structure, $E_{B}$ is the calculated energy of the bulk structure and $A$ is the surface area of the mono-layer. These needed energy data are extracted and calculated by our in-house codes. Electronic band gaps are extracted from the output files of VASP using $pymatgen.io.vasp.outputs$, a function in Python Materials Genomics (Pymatgen).{\cite{jain2011high}} 

\noindent\section{Results and Discussion\label{sec1}}
Our database contains 1,064 EC-MOF structures which are fully relaxed under periodic boundary conditions using DFT calculations. A wide range of data including structural information, formation energies  and electronic properties are extracted and included in our EC-MOF/Phase-I database that can be accessed via https://ec-mof.njit.edu. Such important data is useful for other researchers who need to screen and target attractive structures without additional effort. Here, we discuss these properties in more details.

\noindent\subsection{Structural data analysis}
\begin{figure}[h!]
\centering
\includegraphics[width=0.8\linewidth]{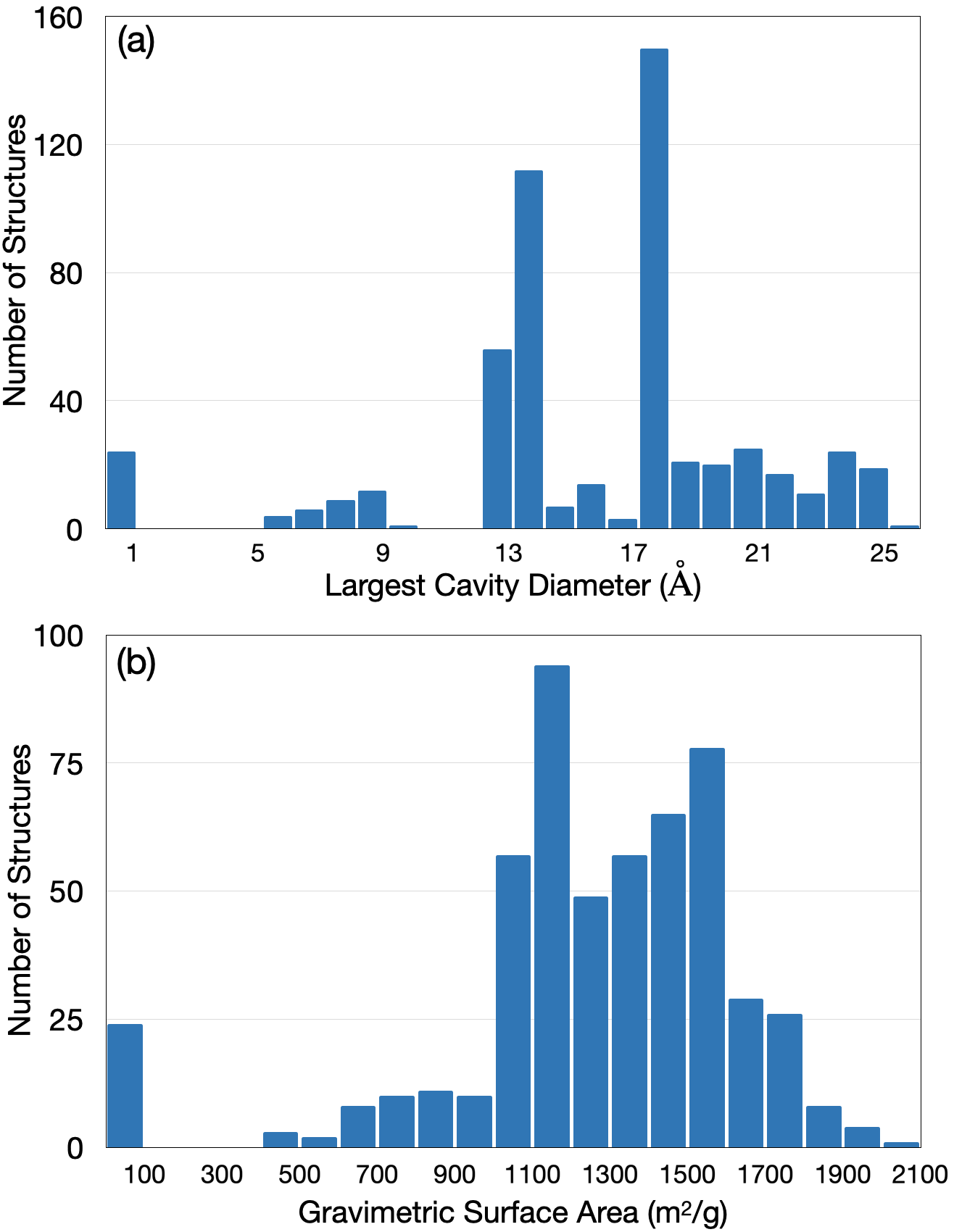}
\caption{Number of structures from our EC-MOF database as a function of the calculated (a) largest cavity diameters and (b) gravimetric surface areas.}
\label{LCD&GSA}
\end{figure}

\begin{figure}[h!]
\centering
\includegraphics[width=0.9\linewidth]{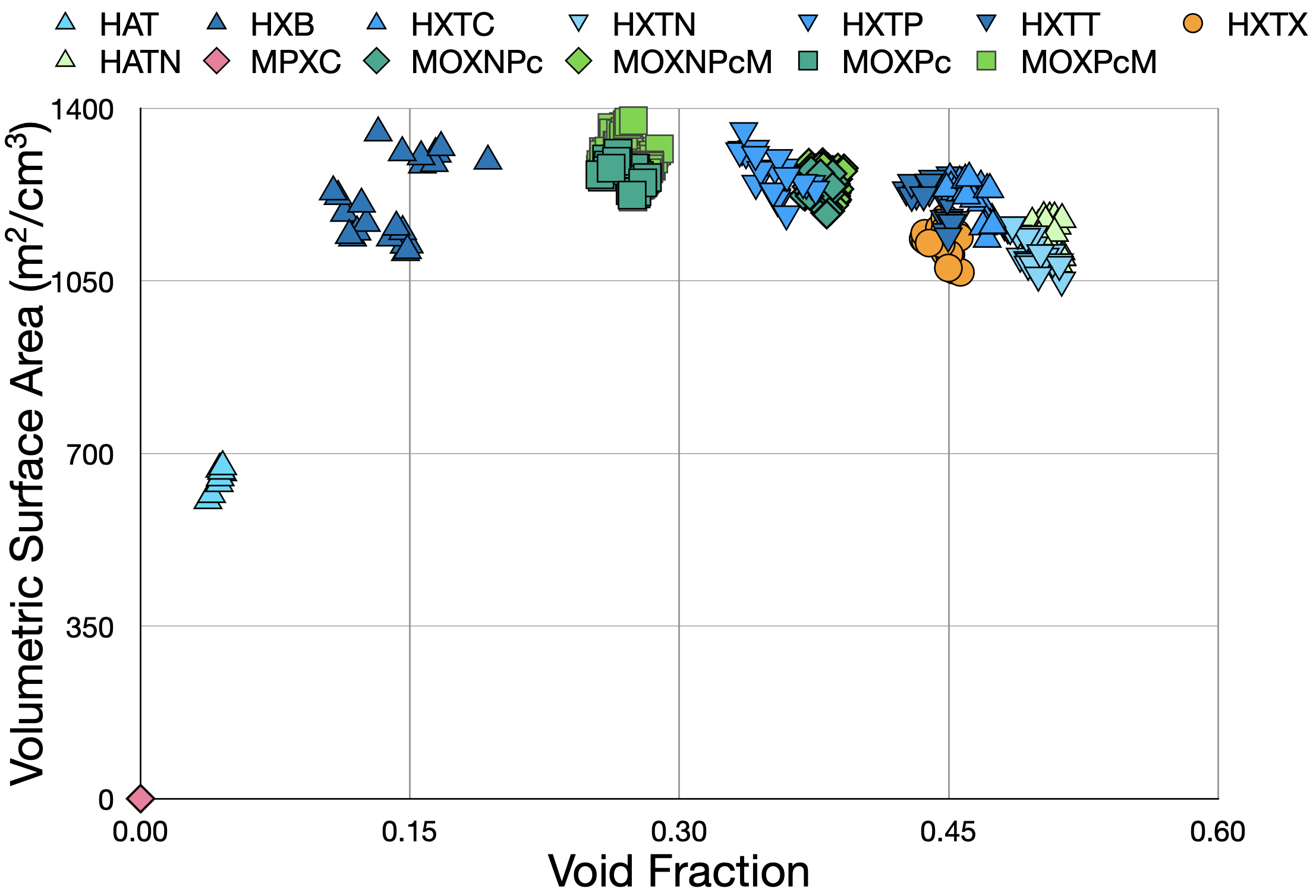}
\caption{Calculated volumetric surface areas as a function of void fraction according to different organic linkers.}
\label{FigureASA}
\end{figure}

During the full relaxation, all 536 bulk EC-MOFs are successfully converged to the force criteria explained before. In case of mono-layer structures, eight MPIC systems failed to maintain their connectivity and topology due to the strong steric hindrance which breaks the bonds between nitrogen and hydrogen atoms. Hence, they were not included neither in the online EC-MOF/Phase-I database nor here. This results in 1,064 structures in EC-MOF/Phase-I database. Another notable point about mono-layer systems is that the optimized structures of HAT-based MOFs are slightly distorted due to steric hindrance between two neighboring organic linkers. Similar phenomena were observed by Lyu \textit{et al.}\cite{lyu2022synthesis} where mono-layer materials of HAT-based MOFs can be only synthesized by an on-surface reaction approach. Here, we first present structural data extracted from 536 optimized bulk systems. Figure~\ref{LCD&GSA}a shows frequency of structures as a function of largest cavity diameter (LCD) which is defined as the diameter of a sphere that can fit into the largest pore of the materials.\cite{zeo++}
 LCD of 2D MOFs in our EC-MOF database ranges from 0 to 25.2 \AA~depending on the size of the organic linkers. Structures with the highest LCD values consist of TC, HATN and TT-based MOFs with the largest linker size of all. Structures with small LCD are mostly C-based MOFs. The high number of connectivities of one C linker to the metal nodes limits the spatial space leading to a small LCD, 0 \AA. It should be noted that 0 value of LCD does not mean that the system is not porous but the diameter of the Helium probe is larger than the diameter of the pore in these structures. The calculated LCD of the rest of the structures is qualitatively ranked according to the size of their organic linkers. Pc and PcM-based MOFs, where linkers have similar sizes, constitute the big body of MOFs with LCD of $\sim$13 \AA, Figure~\ref{LCD&GSA}a. Similarly, NPc and NPcM-based MOFs constitute the peak at $\sim$17 \AA~in Figure~\ref{LCD&GSA}a. Figure~\ref{LCD&GSA}b shows frequency of structures as a function of gravimetric surface area (GSA). In the range above 1700 $m^2/g$, TC, HATN, TN and TT linkers along with several tetragonal linkers take the lead in inducing the largest GSA. Due to the unique structure of the TC linker, structures of TC-based MOFs have not only conventional 1D channels but another 1D channel with a smaller radius due to the porosity inside the linkers. Therefore, TC-based MOFs possess the highest GSA with an average value of 1711 $m^2/g$ over all combinations of metals and functional groups, as reported in Table~\ref{TableAVE}. 
 \begin{table}[]
\begin{tabular}{ccc}
\hline
{\color[HTML]{262626} \textbf{Linker}} & {\color[HTML]{262626} \textbf{LCD (\AA)}} & {\color[HTML]{262626} \textbf{GSA (m$^2$/g)}} \\
\hline
{\color[HTML]{262626} TC}&21.359&1711\\
{\color[HTML]{262626} HATN}&23.800&1639\\
{\color[HTML]{262626} TN}&23.908&1589\\
{\color[HTML]{262626} TT}&19.737&1502\\
{\color[HTML]{262626} NPc}&17.637&1438\\
{\color[HTML]{262626} NPcM}&17.659&1434\\
{\color[HTML]{262626} TX}&20.444&1366\\
{\color[HTML]{262626} TP}&15.381&1201\\
{\color[HTML]{262626} Pc}&12.962&1120\\
{\color[HTML]{262626} PcM}&12.990&1105\\
{\color[HTML]{262626} B}&7.866&663\\
{\color[HTML]{262626} HAT}&5.911&313\\
{\color[HTML]{262626} C}&0&0\\                   
\hline
\end{tabular}
\caption{\label{TableAVE}Average largest cavity diameter and gravimetric surface area values with respect to each organic linker.}
\end{table}
 Similar result was reported in the work by Park \textit{et al.} \cite{pham2022imparting}, where CuHHTC was synthesized and featured as one special 2D MOF with enhanced surface area. In their experimental work, an unprecedentedly high GSA of up to 1196 $m^2/g$ was measured that can be compared to the CuHHTC within our EC-MOF database with a GSA of 1737 $m^2/g$. GSA of TT-based 2D MOFs is also reported by Dinc\u a \textit{et al.} \cite{dou2021atomically} where the GSA of the synthesized CuHHTT is around 1360$\pm$20~m$^2/g$ compared to a value of 1549 m$^2$/g calculated in this work. According to our average calculated GSA data, TT-based 2D MOFs rank in the 4$^{th}$ place as shown in Table~\ref{TableAVE}. Structures found in the lowest range of GSA, i.e., below 500 $m^2/g$, are mainly C and HAT-based MOFs due to the specific connection between organic linkers and metal nodes of these MOFs and the size of probe used in Zeo++. The practical implication of these small cavities will be adsorptive separation of smaller size guest molecules for application as molecular sieves. When comparing our GSA data in EC-MOF database and experimentally measured surface area, our calculated GSA values are always higher than what is measured experimentally by
Brunauer–Emmett–Teller (BET) equations. The discrepancy can be mainly related to (1) the size of crystal particles compared to prefect bulk crystals in EC-MOF database, (2) presence of defects in synthesized materials, and (3) pores that are not completely evacuated from solvent molecules. Notably, the GSA data in our EC-MOF database qualitatively shows similar trends with available experimental results. We note that our GSA data provides the maximum surface area that each EC-MOF can possibly reach regardless of the synthesis and activation procedure. To reveal the relation between spatial space and accessible surface area, we converted GSA to volumetric surface area (VSA) and plotted them against void fraction of 13 different organic linkers as shown in Figure~\ref{FigureASA}. 
Each linker is highlighted by distinct color and shape. In agreement with GSA data, MOFs containing TN, HATN, TC, TX and TT linkers possess high void fraction ranging from 0.43 to 0.51 due to the size of these linkers. The data points located between 0.1 and 0.4 void fraction also have high VSA values. However, the difference between VSA values of MOFs with different linkers in this region is less noticeable compared to the difference in their GSA values, as shown in Table~\ref{TableAVE}, due to inclusion of density of MOFs in converting GSA to VSA. The void fraction of MOFs with C and HAT linkers are as low as 0.07. Except these two classes, the VSA of the rest of structures are all higher than 1000 $m^2/cm^3$ while their void fraction ranges all the way from 0.1 to 0.5. Hence, regardless of accessible surface area, the appropriate MOFs for adsorption/separation applications should be chosen based on the void fraction as well. For energy/gas storage purposes, gravimetric/volumetric capacities of the active compounds are more relevant to applications than surface area. And they are directly related to GSA and VSA, respectively, according to G{\'o}mez-Gualdr{\'o}n {\textit{et al.}} and Allendorf {\textit{et al.}} \cite{gomez2017understanding,allendorf2018assessment}. It is even possible to calculate the adsorption capacity for specific guest molecules using the surface area data and a correctly fitted equation\cite{gomez2017understanding}. Therefore, our existing surface area data, porosity information, and simulations of adsorption capacity which will be done in the future, are ideal training sets for machine learning algorithms that can predict the adsorption capacity without time-consuming simulations. Furthermore, applications of these MOFs in adsorption/separation should be further investigated by analyzing the guest-host interactions. For instance, maximizing the performance of MOFs in hydrogen storage requires maximizing VSA while preserving a large void fraction.\cite{gomez2017understanding} On the other hand, number of adsorption sites should be the primary concern in case of chemical adsorptions.\cite{li2018novel}

\noindent\subsection{Thermodynamic data analysis}
\begin{figure*}[h!]
\centering
\includegraphics[width=0.99\linewidth]{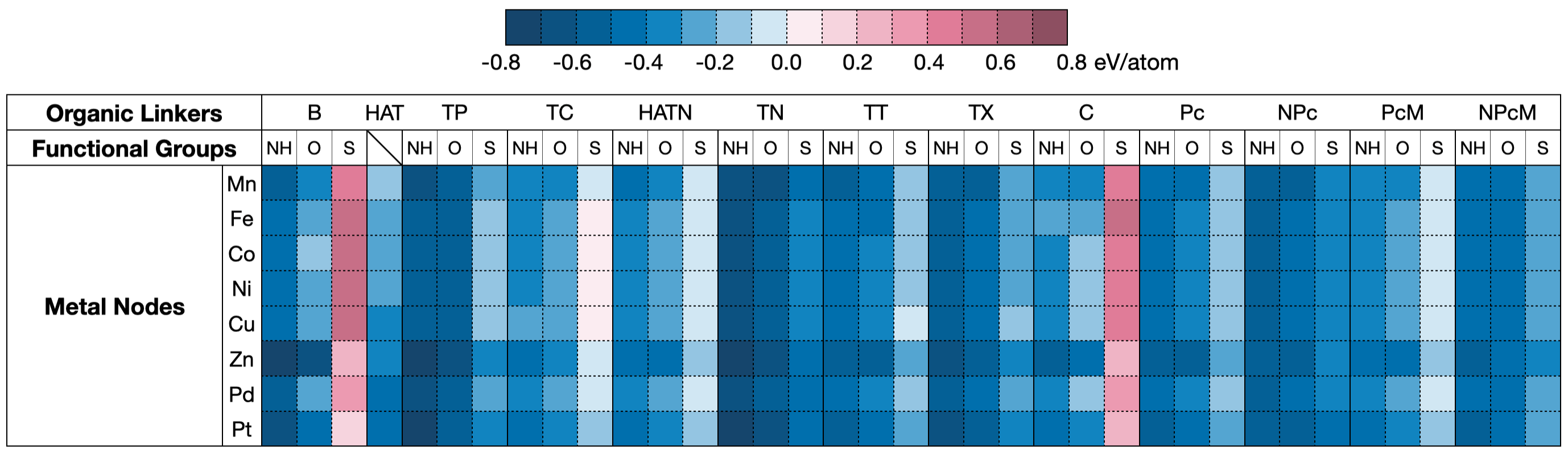}
\caption{Calculated formation energies of all bulk structures contained in our EC-MOF/Phase-I database. The corresponding data for the mono-layers are given in the SI Figure S2.}
\label{FigureEf}
\end{figure*}

Formation energy is one of the most principal parameters to consider when determining whether a hypothetical material will be synthesizable.\cite{rasmussen2015computational, mao2021prediction} Following our high-throughput workflow, calculations of formation energies for 1,064 structures were implemented using Eq.~\ref{eq1} and the results for bulk EC-MOFs are presented in Figure~\ref{FigureEf}. Three mono-layers fail to reach the energy convergence during this stage, which decrease the total number of structures in our EC-MOF/Phase-I database to 1,061.
Calculated formation energies for mono-layer structures showing a similar trend can be found in the SI Figure S2. Both PcM and NPcM-based MOFs which share the same structural building blocks but with different metal centers inside the organic linkers are combined into one data point. This is because the different choices of metal centers were found to have a minor affect on the calculated formation energies, i.e., less than 0.09 eV/atom. Detailed comparisons of these MOFs can be found in the SI Tables S3 and S4. Data points with a blue color in Figure~\ref{FigureEf} mean negative formation energies and show that the corresponding structures are more likely to be synthesizable. The red color on the contrary corresponds to positive formation energies and hence structures that are thermodynamically unstable. Accordingly, 96.1$\%$ of bulk and 92.8$\%$ of mono-layer structures have negative formation energies (see Figure~\ref{FigureEf} and the SI Figure S2). To find a trend in the computed formation energies one should pay attention to the different bonding/interaction motifs that exist in the built MOFs. Overall, in all EC-MOFs there are three main types of bonds/interactions including covalent bonding within the organic motif, coordinative bonding between metals and functional groups of the linkers and comparatively weaker van der Waals interactions between layers. As an example, B, TP and TN linkers could be placed in an incremental sequence considering that each linker is comprised of the previous one plus three more benzene rings. Having more benzene rings in the structure increases the number of covalent bonds which is stronger than the other two interaction/bond types. Hence, the calculated formation energies of TN-based MOFs are more negative than their TP-based counterparts which in turn are more stable than the B-based MOFs, all due to the higher number of covalent bonds. A similar trend can also be observed in the cases of Pc vs. NPc or PcM vs. NPcM where the latter ones have four more benzene rings than the former ones. Figure~\ref{FigureEf} also illustrate the formation energies with respect to three functional groups within each linker family. Its very clear that in each linker family, EC-MOFs with -NH functional groups always possess the most negative formation energies whereas the ones with -S groups possess the least negative or in some cases even positive values. The stability of the coordinative bonds between transition metals and different functional groups is determined by the compatibility between the two interacting units. Nitrogen atoms have lower electronegativity than oxygen leading to a stronger bonding to metals. Compared to sulfur atoms, sizes of nitrogen and oxygen atoms are more similar to carbon atoms, which induces a better overlap between atomic orbitals in the extended $\pi$-conjugated layers. Considering different transition metals, among eight employed metal nodes, Mn$^{2+}$, Zn$^{2+}$ and Pt$^{2+}$-containing MOFs tend to have more negative formation energies regardless of organic linkers and functional groups. Half-filled and fully-filled electronic configurations of Mn$^{2+}$ and Zn$^{2+}$, respectively, are more stable in the same row of periodic table when forming 2+ ions. Other than the standard way of creation of MOFs from different building blocks, physical and chemical properties of the EC-MOF based layered materials can be further modulated by tuning their inter-layer interactions by reducing the number of stacked layers all the way down to a mono-layer. Consequently, mono-layer materials exfoliated from their corresponding bulk systems may possess unconventional properties in the fields of gas adsorption\cite{liu2021fabrication}, optics\cite{zheng2020recent} and electrocatalysis\cite{khan2022current}. EC-MOF mono-layers can be synthesized through top-down or bottom-up approaches. Chemical exfoliation methods such as intercalation and electrochemical exfoliation are among the most common techniques in the top-down strategy.\cite{zheng2020recent} On the contrary, bottom-up approach is a more efficient and convenient approach to build mono-layer materials by choosing ideal precursors, modulators and surfactants.\cite{khan2022current} To shed light on the possibility of reaching at a mono-layer EC-MOF, we have calculated the inter-layer binding energies (E$_b$) according to Eq.~\ref{eq2} for all structures gathered in EC-MOF/Phase-I database. 
\begin{figure}[h!]
\centering
\includegraphics[width=0.8\linewidth]{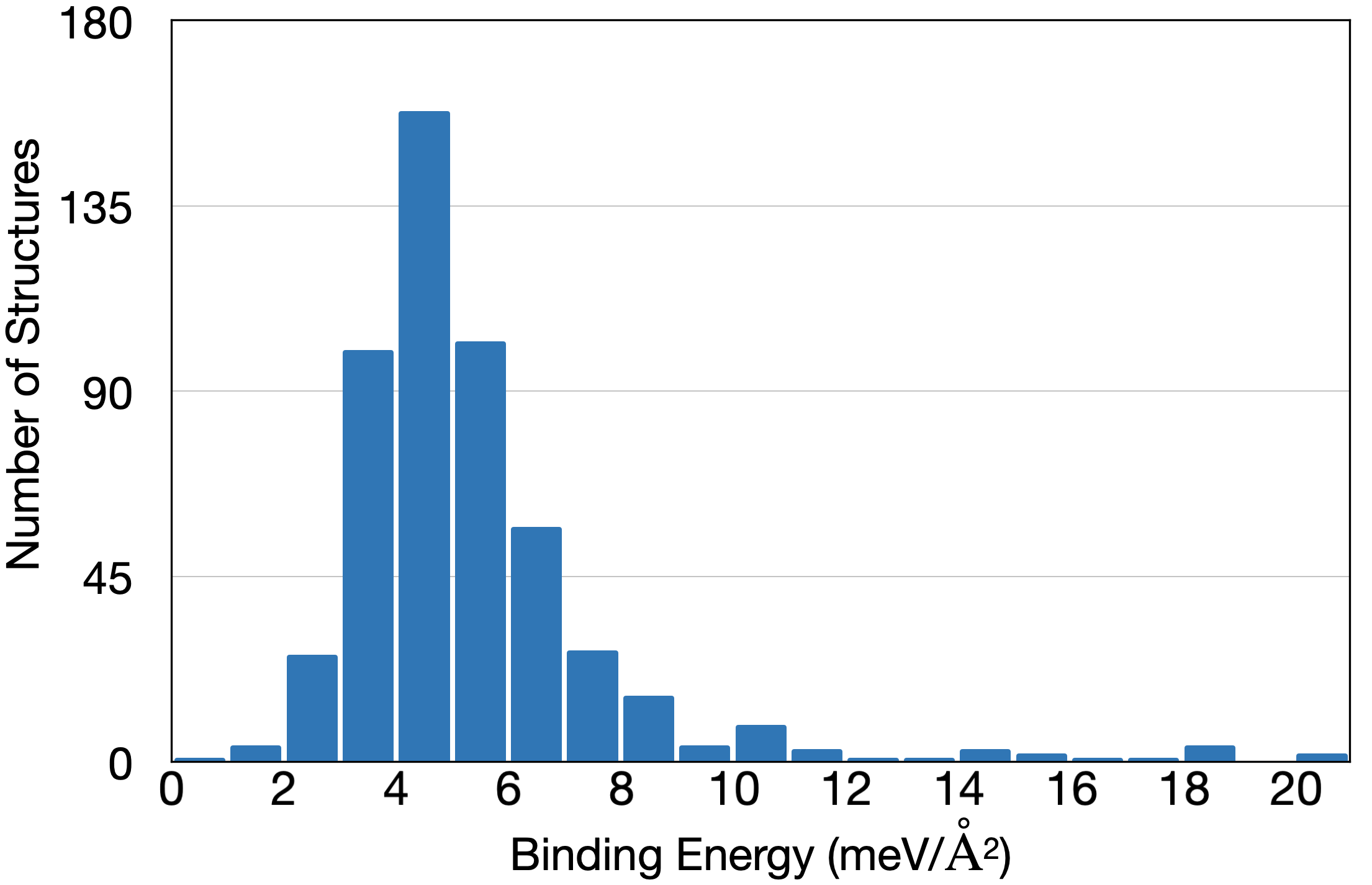}
\caption{Distribution of structures according to their inter-layer binding energies.}
\label{FigureEb}
\end{figure}
Figure~\ref{FigureEb} shows the histogram of the distribution of E$_b$ in the database with the prominent range of 0 to 20 meV/\AA$^2$ with a bin width of 1 meV/\AA$^2$. The E$_b$ of the majority of our EC-MOFs lie in the range of 2-9 meV/\AA$^2$. This can be compared to the interlayer binding and exfoliation energies for a large number of layered compounds, including graphite and MoS$_2$, which are around 20 meV/\AA$^2$ and are considered as candidates for a successful exfoliation.\cite{Nieminen:2012} It should be mentioned that the low E$_b$ values of EC-MOFs in comparison to other layered materials relates to the high porosity of our structures. This indicates the promising potential of most of these EC-MOFs for thin-film fabrication which is of utmost importance in compact device implementation. It is also worthwhile noting that, according to Eq. (2), calculated E$_b$ values are inversely proportional to the surface area of the mono-layer. Hence, within a family of porous materials one will naturally obtain smaller E$_b$ values for the systems with bigger pore sizes in their unit cells than the ones with smaller pore sizes, given similar inter-layer van der Waals interactions. For example, EC-MOFs with calculated E$_b$ values larger than 13 meV/\AA$^2$ are mainly HAT and C-based MOFs with possibly strong $\pi-\pi$ interactions per unit area in addition to having some of the smallest pore sizes/surface areas. To be noted, two structures, MnHAT and PtHHTP show E$_b$ values higher than 20 meV/\AA$^2$, 23.17 and 21.57 meV/\AA$^2$, respectively. 

\noindent\subsection{Electronic property data analysis}
\begin{figure}[t!]
\centering
\includegraphics[width=0.99\linewidth]{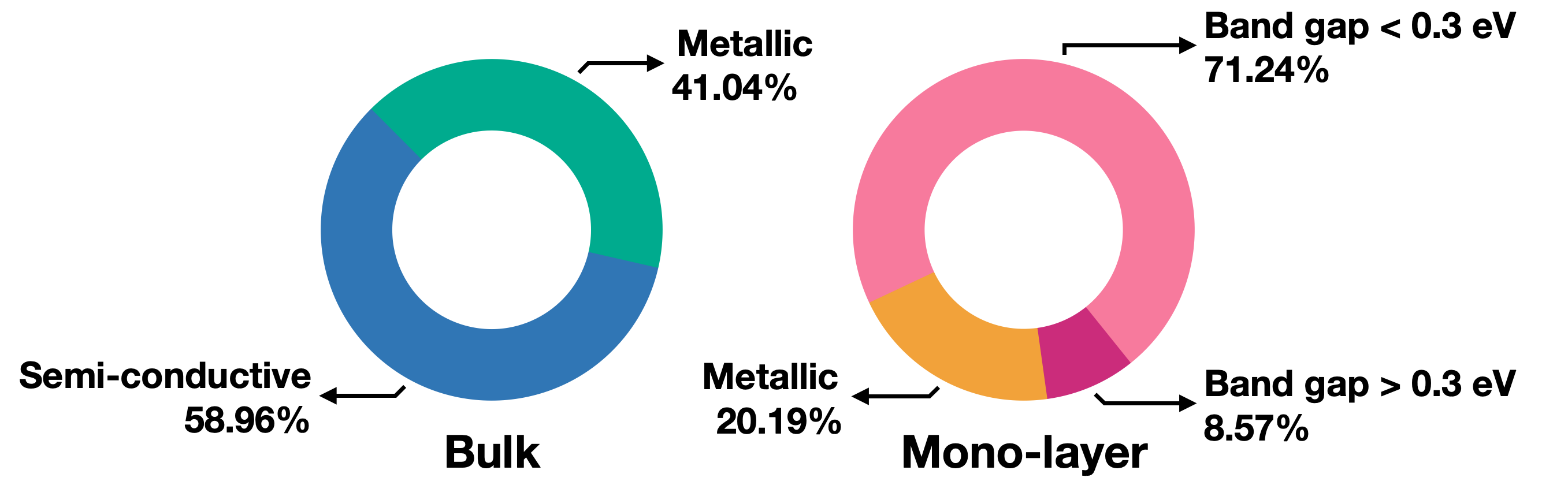}
\caption{Distribution of metallic and semiconductor systems in bulk (left) and mono-layer (right) materials. Mono-layer semiconductors are further divided into 2 regions according to their band gap values.}
\label{FigureConductiveDist}
\end{figure}

\begin{figure*}[h!]
\centering
\includegraphics[width=0.99\linewidth]{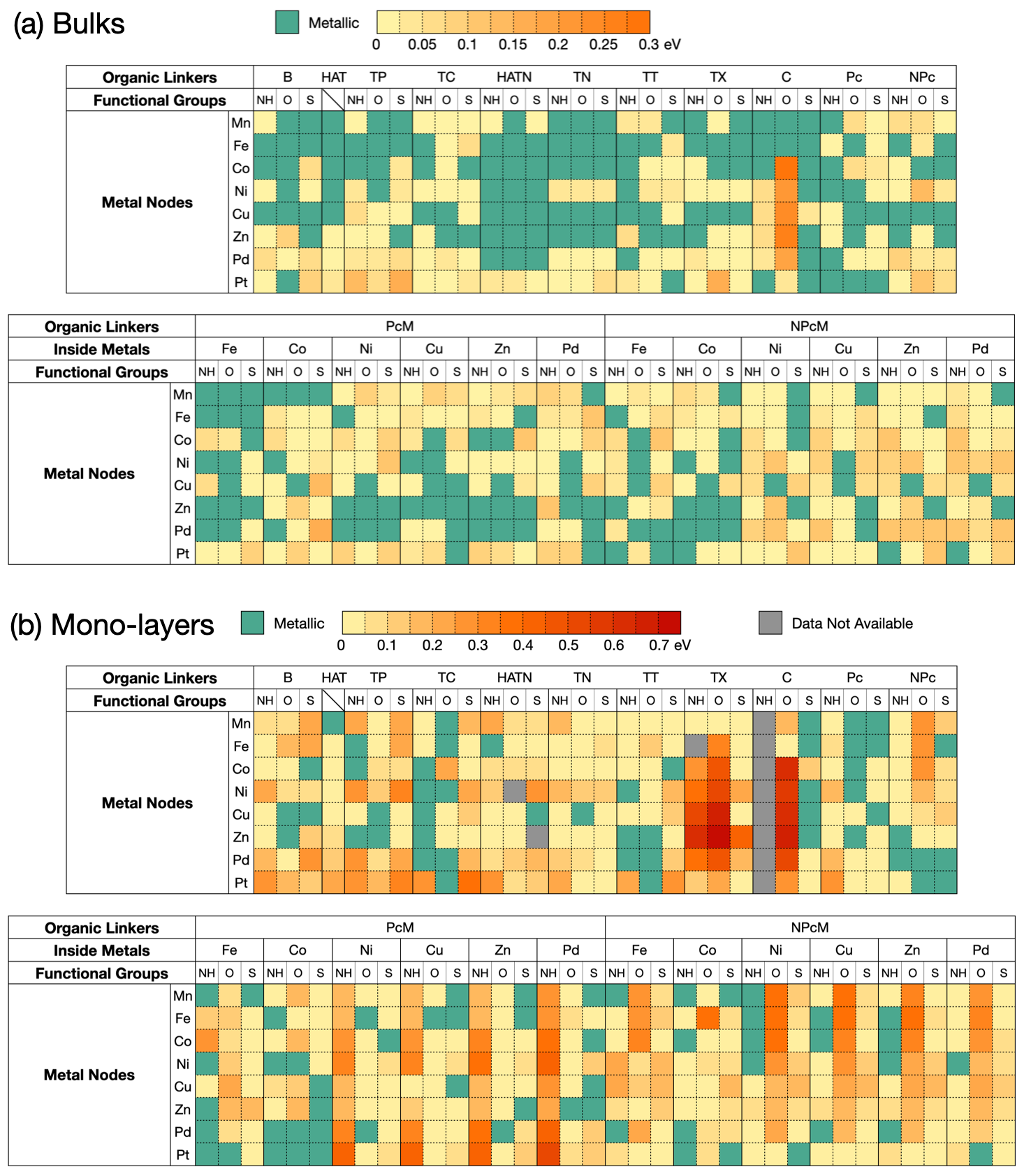}
\caption{Calculated band gaps of (a) bulk and (b) mono-layer systems.}
\label{FigureBandGap}
\end{figure*}

\begin{figure}[ht!]
\centering
\includegraphics[width=0.99\linewidth]{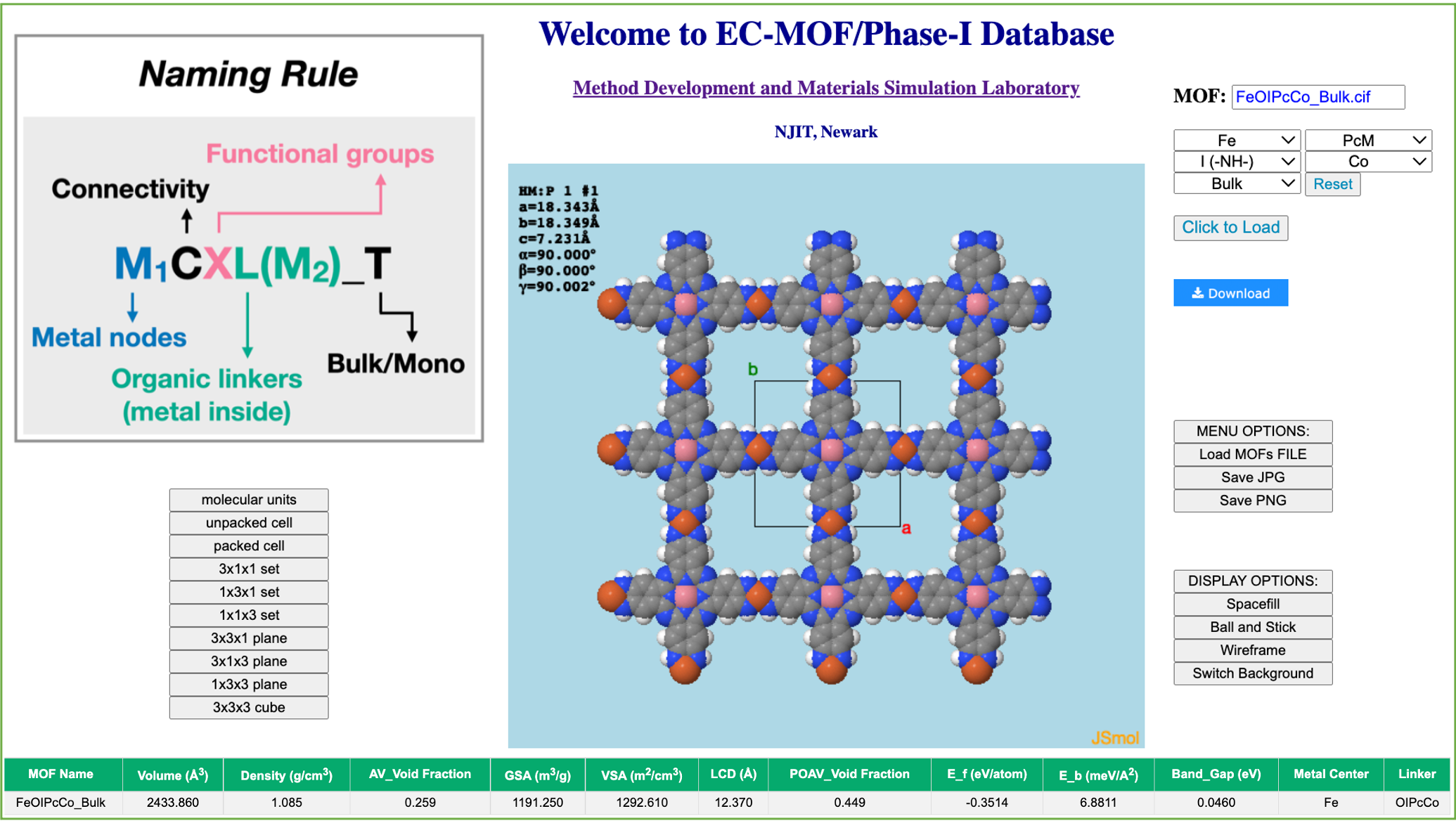}
\caption{A screenshot of the graphical user interface developed for the EC-MOF/Phase-I database (https://ec-mof.njit.edu).}
\label{FigureDatabase}
\end{figure}

The electrically conductive behavior of $\pi$-stacked layered MOFs is in contrast to other conventional MOFs that are mostly classified as insulators where electrons are highly localized within the framework.\cite{kung2019electronically} To provide more insights on the electrical conductive behavior of the structures gathered in the EC-MOF database, we have calculated their fundamental electronic band gaps as explained in section 2. As a result, 40.9$\%$ of bulk EC-MOFs are calculated to be metallic compared to 20.2$\%$ for mono-layers, Figure~\ref{FigureConductiveDist}. This can be explained by the absence of inter-layer charge transport pathway along the stacking direction for the mono-layer structures. We also calculated the ratio of semiconductors within each class of EC-MOFs with different organic linkers, see SI Table S5, and found that for most structures, the ratio of semiconductors in mono-layers are indeed higher than the ratios in bulk systems.
Figure~\ref{FigureBandGap} shows the distribution of conductive behaviors of all structures in the EC-MOF/Phase-I database. 325 structures are found to be metallic with 219 of them being bulk and the rest of 106 being mono-layer systems. For the semiconductors, the maximum band gap value of 0.744 eV is reached by ZnHHTX\_Mono. Among all 536 bulk systems, band gaps range from 0.001 eV to 0.294 eV where the maximum band gap is reached by CoPHC\_Bulk. In Figure~\ref{FigureBandGap}(a), we can find that most bulk materials are metallic or semiconductor with a narrow gap that is smaller than 0.15 eV. According to our previous work\cite{zhang2021metal}, a transition of conductive behavior between metallic and semiconductor can be easily induced by temperature, pressure or solvent because of the intrinsic flexibility of these materials. Therefore, such conductive behaviors indicate the great advantage of EC-MOFs compared to conventional MOFs. 
Indeed, some of them have already been proven to be promising candidates for applications in batteries\cite{cai2019lithium}, photodetectors\cite{arora2020demonstration} and voltammetric detection\cite{ko2020employing}. In Figure~\ref{FigureBandGap}(b), not only the number of semiconductors increases in mono-layers but the band gap values also increase. As can be seen from this Figure, red data points gather under the TX and C linkers. All of TX-based mono-layers are found to be semiconductors with large band gaps. By comparison, only 13 out of 24 TX-based bulk systems are semiconductors with the maximum gap of 0.157 eV found in the database. This can be explained by the disconnection of $\pi$-conjugated benzene rings of TX linker, so the inter-layer conducting is more dominant in such systems. Because PHC-based MOFs have the highest number of connectivity, 12 hydroxyl groups per linker, the most electronegative oxygen atom tend to localize the charge from $\pi$-system in C linker, which leads to a wider gap in these systems. 
Another interesting trend is found in PcM and NPcM-based mono-layers with Ni, Cu, Zn and Pd inside metals. These mono-layers with imino functional groups always possess high band gap values because of the presence of inside metals compared to Pc and NPc-based mono-layers. However, no trend was found for the Fe and Co as inside metals. Therefore there could be some long range interactions at play between different metals and functional groups that needs to be further investigated. Finally, an exclusive graphical user interface was developed for our EC-MOF/Phase-I database (Figure~\ref{FigureDatabase}) and made available to the community and public (https://ec-mof.njit.edu) where all the curated structures at the DFT level can be visualized and downloaded with their relevant calculated properties tabulated.

\noindent\section{Future Work\label{sec1}}
Our EC-MOF/Phase-I database provides an easily accessible computationally ready data bank for $\pi$-stacked layered electrically conductive MOFs with diverse information on their structural and electronic properties.
We plan to update the database as the research in this field grows and expands. It is necessary to update this database in two directions, diversity of the MOFs included and types of properties calculated for them. As stated previously, we have elected to restrict the first version of the database according to structural features that induce the highest electrical conductivity. However, as mentioned before, this is an active area of research and new organic linkers are introduced almost everyday.\cite{guo2022dialytic} Also, MOFs with various topologies will be included in the future versions of the database, e.g., three-dimensional MOFs\cite{sun2017iron, darago2015electronic}. Based on our building strategy, permutations of building blocks will give a comprehensive understanding of these classes of MOFs as well. On the other hand, other relevant properties need to be calculated, such as performance on adsorption/storage of common gas molecules\cite{colon2017topologically}, partial atomic charges normally used to interpret trends while modeling chemical reactions\cite{rosen2022high} and density of states/band structures which reveal detailed charge transport pathways. 
Recently, machine learning techniques have shown great promise in materials science research for prediction of formation energies\cite{gong2021calibrating}, adsorption energies\cite{panapitiya2018machine, wang2020electric}, band gap values\cite{rosen2021machine} and designing new materials\cite{mai2022machine}. The created crystal structures and their calculated property data gathered in our EC-MOF/Phase-I database provide an ideal data set for applying various machine learning techniques in order to explore their potentials for different applications. 

\noindent\section{Conclusion\label{sec1}}
In this work, we introduced for the first time an exclusive database for electrically conductive MOFs containing computationally-ready structures and their property data. A total number of 1,072 structures are created by taking permutations among the subsets of different structural building blocks using our in-house package, Crystal Structure Producer (CrySP). Multiple stages of calculations are applied to the database by applying a high-throughput screening workflow to optimize the structures and calculate their different property data. 1,064 out of 1,072 structures were successfully optimized at the DFT level and 1,061 of them successfully completed all stages of calculations whose properties including largest cavity diameter, gravimetric/volumetric surface area, void fraction, formation energy, inter-layer binding energy and electronic band gap added into the database. Obtained trends for different classes of MOFs were discussed in great details and different materials were classified and analyzed according to their structural or electronic properties providing comprehensive and important information on different families of EC-MOFs. Finally, an exclusive graphical user interface was developed and released for this database where all curated structures at the DFT level can be visualized and downloaded with their relevant calculated properties tabulated.

\noindent\section{Supporting Information}
Mathematical equations for calculation of cell parameters in CrySP,
Hubbard U parameters used in this work,
chemical potentials calculated in VASP using the most stable configurations of the pure elements,
calculated formation energies, and computed ratios of semiconductor materials to total number of structures divided between different organic linkers.

\noindent\section{Author Information}
\noindent\subsection{$^{\dagger}$Present Address:}

\noindent Division of Energy, Matter and Systems, University of Missouri $-$ Kansas City, Kansas City, MO 64110, United States, mmomenitaheri@umkc.edu (M.R.M.)

\noindent\section{Acknowledgments}
\noindent FAS acknowledges support from the start-up fund provided by New Jersey Institute of Technology (NJIT). This work used Bridges2 at Pittsburgh Supercomputing Center through allocation CHE200007 from the Extreme Science and Engineering Discovery Environment (XSEDE),\cite{xsede} which was supported by National Science Foundation grant number 1548562. This research has been (partially) enabled by the use of computing resources and technical support provided by the HPC center at NJIT.

\bibliography{bib}
\bibliographystyle{achemso}


\end{document}